\documentclass[aps,preprint,superscriptaddress,floatfix]{revtex4}

\usepackage[pdftex]{graphicx}
\usepackage{dcolumn}
\usepackage{bm}
\usepackage{amssymb}
\usepackage[utf8]{inputenc}
\usepackage{color} 
\usepackage{soul}  

\begin{document}

\title{Parallel-Tempering Monte-Carlo Simulation with Feedback-Optimized Algorithm  
Applied to a Coil-to-Globule Transition of a Lattice Homopolymer
}

\author{K. Lewandowski}
\author{P. Knycha{\l}a}
\author{M. Banaszak}
\email[]{mbanasz@amu.edu.pl}
\homepage[]{http://www.simgroup.amu.edu.pl}
\affiliation{ Faculty of Physics,
A. Mickiewicz University \\
ul. Umultowska 85,
61-614 Poznan,
Poland
}

\date{\today}

\begin{abstract}
We present a study of the parallel tempering (replica exchange) Monte Carlo method, with special focus on the feedback-optimized parallel tempering algorithm, used for generating an optimal  set of simulation temperatures. This method is applied to a lattice simulation of a homopolymer chain undergoing a
coil-to-globule transition upon cooling. We select  the optimal number of replicas for different chain lengths, N = 25, 50 and 75, using 
 replica's round-trip time in temperature space, in order to determine energy, specific heat, and
squared end-to-end distance of the hopolymer chain for the selected temperatures. 
We also evaluate  relative merits of this optimization method.

\end{abstract}


\maketitle

\section{Introduction}
Monte Carlo (MC) simulations are widely used in polymer  modeling \cite{binder08} in order to extract the relevant thermodynamic and structural properties. The MC method is based on generating a random set of points  in configuration space $x_1, x_2, ..., x_n$ (where $n$ is a large number) which resemble the distribution of such points in thermal equilibrium. The practical problem is to limit $n$, also referred to as the number of MC steps (MCS), to a manageable size, which would result in realistic times of simulation (we define MCS step as a one attempt to move each particle in a simulated system). For example, the standard Metropolis Algorithm (MA)  \cite{binder08} provides a straightforward way to generate such points, but the number of MCS to reproduce the equilibrium distribution is often prohibitively large. While the MA works quite well for high temperatures (typically above a phase transitions), it tends to generate  traps in the local free energy minima for low temperatures (typically below a phase transition). Those  traps often result in  unreliable estimates of the sampled properties. To solve this problem,  many modifications of the MA are proposed \cite{Wang-Landau, cerny, kirkpatrick}. One of those  is parallel tempering (PT) method \cite{swendsen_wang, earl}, in which by parallel simulation of many replicas, in the relevant temperature range, the  energy barriers can be overcome. In general, one simulates $M$ copies of a system for a set of temperatures, $T_1 < T_2 < \dots < T_{M-1} < T_M$, where $i=1, 2, \dots, M$.  The minimum and the maximum temperatures, $T_1$ and $T_M$,  are fixed and the intermediate temperatures are selected to provide an optimal walk of replicas in the  temperature space. The adjacent replicas are exchanged, at a given frequency in MCS, in random order with the following probability:
\begin{equation}
p(T_i^* \leftrightarrow T_{i+1}^*) = \min[1, \exp(-(\beta_i - \beta_{i+1})(U_{i+1} - U_i))]
\end{equation}
where $\beta_i = 1 / k_BT_i^*$, $k_B$ is the Boltzmann constant,  and $U_i$ is potential energy of replica at $T_i^*$. If the system  is trapped at  lower temperature, it can be heated up in higher temperature and overcome the energy barrier. By using such scheme, one can get better statistics in less MCS, but one is required to simulate $M$ replicas in parallel. However, to run such simulation one has to set some parameters: number of replicas (temperatures) $M$,  temperature set, $ \{T_1, T_2, ..., T_M\}$,  and exchange interval, $S$. It is known that quality of the PT method strongly depends on those parameters, and if they are not chosen correctly, simulation can give the same results as without the PT. It is not clear how many replicas are optimal, and it is hard to establish the appropriate criteria for this choice. One approach is to consider some thermodynamic quantity (for example the specific heat, $C_v$) and to test it how much CPU-time  is needed to obtain correct result,  or run many simulations with the same number of MCS (or total MCS of all replicas),  and to test measurement accuracy for  this quantity.   In another approach, one can find the optimum $T^*$ set that gives the shortest round-trip times of replicas in temperature space.

Feedback-optimized PT (FOPT) algorithm, which we use in this work,  is designed to find temperature sets that maximize the current of replicas from the lowest temperature to the highest one (details in reference \cite{feedback}). In this algorithm we mark each replica with a flag  $up$ (when it reaches $T_1^*$) or $down$ (when it reaches $T_M^*$).  Initially the  replicas do not have any flag, but they acquire one of them if any of the above conditions is met. During the simulation,  we generate histograms $f_{up}(T_i^*)$ and $f_{down}(T_i^*)$.  Before exchanging replicas, we check all replicas and increment histograms in such way that $f_{up}(T_i^*)$ is increased if  replica with  $T_i^*$ has an  $up$ flag, and   $f_{down}(T_i^*)$ is increased if it has  a  $down$ flag. If replica does not have any flag then neithter of histograms is increased. From this data we create  the $f(T_i^*)$ histogram:

\begin{equation}
f(T_i^*) = \frac{f_{up}(T_i^*)}{f_{up}(T_i^*)+f_{down}(T_i^*)}
\end{equation}

The FOPT uses $f(T_i^*)$ histogram to optimize temperature set in order to minimize round-trip times. 

In this work, we use round-trip time to find optimal parameters set. We test systems of different sizes, $N$, and with different number of replicas, $M$,  to find the optimum parameter set. We do not vary the exchange interval, $S$, which is set to $200$ MCS. In the future, we intend to  test  the $S$ effect on the  PT simulation results.

It is worth to notice,  that there are also other methods for  generating optimal temperature sets \cite{gront_kolinski, constant_entropy_model, predescu}. One of them, is proposed by Sabe \textit{et al.} \cite{constant_entropy_model}, and is based mainly on the work of Kofke \cite{kofke} and Kone and Kofke \cite{kone_and_kofke}, in which the optimal temperature sets  are found by using  the condition of constant entropy increase for the adjacent replicas. 
As far as we know, round-trip time of replicas with FOPT was not used as a criterion for finding optimal number of replicas in PT simulation. 

\section{Model}
The simulation is performed  on the face centered cubic (FCC) lattice with coordination number $z = 12$ and the bond length $l = a\sqrt{2}$ where $a$ is the FCC lattice constant. Chain bond are not allowed to be stretched or broken, and the usual periodic conditions are imposed. Lattice sites, which do not have the  chain segments,  are considered to contain the implicit solvent. The interaction is limited to the nearest neighbors ($z = 12$), and the interaction parameter, $\epsilon $ is related to the Flory $\chi$ parameter by the following equation:

\begin{equation}
\chi = \frac{(z-2)\epsilon}{k_BT}
\end{equation}

This parameter, $\epsilon$, serves also as an energy unit to define the reduced energy per segment, and the reduced temperature as

\begin{equation}
\begin{array}{c}
E^*/N = \left(\frac{E}{\epsilon}\right)/N, \\
T^* = k_BT/\epsilon
\end{array}
\end{equation}
where, as before,  $N$ is the number of chain segments. 

In this  simulation we use a single homopolymer chain with lengths $N = 25$, $50$ and $75$ segments on the  FCC $30 \times 30 \times30$ lattice. Elementary moves consists of rotation of a segment inside a chain, moving first and last segment around neighbouring segment and slithering-snake move.
Detailed description of this  model can be found in references \cite{fcc_simulation_model, fcc_offlatice}.

\section{Results and discussion}
\subsection{Simulation}

First, the system is equilibrated in athermal limit ($\epsilon/(k_BT) = 0 $), then we perform $2.5\times  10^5$ MCS to equilibrate system in the specified  temperature range,  and another $2.5\times  10^5$ MCS is used to sample data. During the simulation we collect data for $f(T_i^*)$ histogram and calculate the number of round trips that each replica performs. To get better $f(T_i^*)$'s we assume that the minimum number of round-trips for each replica should be $4$. If this requirement is not met, we extend the simulation ``time''  by another $5\times  10^5$ MCS. For each polymer length we perform many simulations with different number of replicas. For $N = 25$ we use $M = 6, 8, 10, 12, 14, 16, 20, 24, 28, 32, 40, 50, 60, 70, 80$ and $94$; for $N = 50$ we use $M = 8, 10, 12, 14, 16, 18, 20, 22, 24, 26, 28, 30, 34, 38, 42, 46, 50, 54, 58, 62, 66, 70, 74, 78, 86$ and $94$; for $N = 75$ we use $M = 10, 12, 14, 16, 18, 20, 22, 24, 26, 28, 30, 38, 46, 54, 62, 70, 78, 86$ and $94$. All systems are simulated in temperatures range  $T_1^* = 1$ to $T_M^* = 15$.

In each simulation we perform 6 rounds of the FOPT, starting with linear $T^*$ set chosen for the first round. \hl{We choose 6 iteration arbitrarily. As one can see in Figure }\ref{temperatures_set}\hl{ temperature set in round 4 and 5 is very similar and if we run more rounds, we will get similar sets fluctuating around ideal one.} In the last round we run $10^7$ MCS to collect data to find optimal number of replicas. First $5\times 10^6$ MCS are used to equilibrate system and another $5\times 10^6$ MCS to collect the data (for longer chains and greater number of replicas we doubled or even tripled the number of  MCS to obtain better statistics).

Before we show the results for optimal number of replicas, we  look into the simulation process from the 1st round to the 5th round of the FOPT for chain length $N=75$ and $M=14$ and  $30$ (the second $M$ is used as a reference for simulation with $M=14$). In Figure \ref{trace} we present three traces of selected replicas in the $T^*$ space from  the 1st round. As one can see each of the presented replicas reaches $T_1^*=1$ and $T_M^*=15$ many times, but  more often it reaches higher temperatures.

To use the full potential of  the PT method, the $T^*$ set  should be optimized  in such way that each replica spends about the same amount of time at  each temperature. We use the FOPT algorithm for that goal.
Figure \ref{updown_histogram} presents histogram $f(T_i^*)$, Figure \ref{temperatures_set} presents generated $T^*$ set for each simulation round and Figure \ref{probability} presents exchange probability between adjacent replicas in the 1st, the 2nd and the 5th round. In the 1st round we use linear $T^*$ set, in next rounds temperatures are generated from the  $f(T_i^*)$ histogram obtained  in the previous round. The 2nd round yields a reasonably optimized  $T^*$ set, and in the next rounds, the $T^*$ set gets closer to the optimal solution. Black line in Figure \ref{updown_histogram} indicates ideal behavior. In the first round we can see that the $T_1^*$-replica  is trapped and that its exchange probability with the  $T_2^*$-replica is about $1\%$. Exchange probabilities for other replicas are much higher and they grow with increasing $T^*$. In the 2nd round the bottleneck at $T_1^*$ is removed and there is a significant  improvement in the $f(T_i^*)$ histogram, but we can notice another bottleneck at $T_9^*$ with the exchange probability of about $9\%$. This bottleneck is removed in the following rounds.
As seen from exchange probability for the 5th round, the optimal solution has varying exchange probabilities. In general, they should be higher for lower temperatures and for $T_i^*$ with larger specific heat.

\subsection{The long chain properties}
Now we  focus on selected thermodynamic and structural properties of the polymer chain of length $N = 75$ with $M=14$ replicas. It is well known that a single homopolymer chain undergoes coil-to-globule transition upon cooling (in general, upon decreasing the solvent quality). In Figure \ref{en} we present reduced energy per segment, in Figure \ref{cv} the specific heat, and in Figure \ref{r2} the squared end-to-end distance for 1st round (dashed), for the 5th round (dotted), and for the 6th round after $2\times 10^7$ MCS (solid) that is treated as a reference. From $T^* = 15$ to $7.2$ the chain is in the swollen coiled state, then at about $T^* \approx 7.2$ it undergoes coil-to-globule transition. For lower $T^*$ it is in a globular state, but at about $T^* \approx 1.4$ there is another ordering effect, probably crystallization on the FCC network. It corresponds to forming an optimal structure within used lattice. As seen in Figure \ref{ordering}, globule below this transition is highly ordered.  As seen in Figure \ref{en},  the energy for each round is almost the same, but the specific heat is not. Since in the 1st round we use  linear $T^*$ set, we miss the low temperature ordering effect at about $T^* \approx 1.4$. Because the $T_1^*$-replica is most likely  trapped in a free energy minimum, probing specific heat for $T_1^*$ gives a wrong result. But with optimized $T^*$ set for the 5th round, we observe  a considerable improvemanet of the specific heat for the same number of MCS. 

As we can see, the  PT method without optimization  also gives reasonable results.  It would be, therefore,  reasonable to compare simulation results from simulation with and without the PT. Such comparisons can be found, for example,  in reference \cite{sikorski_star_branched_polymers}.
It is worth to notice that the FOPT not only speeds up the simulation process (giving better sampled data approximation) but also it sets temperatures in such way that the relevant temperature ranges are sampled more accurately (by arranging temperatures in areas where the specific heat is higher). 

\subsection{Optimal number of replicas}
To find optimal number of replicas for each system size, we perform simulations with different number of replicas and in each case we measure the round-trip time (to get from $T^*_1$ to $T^*_M$, and then back from $T^*_M$ to $T^*_1$).
Figure \ref{round_trip_time_not_normalized} presents average round-trip time for polymers length N = 25 (solid line), 50 (dashed line) and 75 (dotted line). For each N, the smallest possible number  of replicas is chosen to run the PT so that the round-trip is actually performed. For smaller M's,  the replicas cannot perform any round-trips in the  temperature space. However if we had a better $T^*$ set in the 1st round, it might  run correctly. As we can see in Figure \ref{round_trip_time_not_normalized} when we increase the chain length, the replicas need more MCS to perform one 
round-trip. For $N=25$,  $50$ and $75$ the smallest round-trip time is  $21 \times 10^3$, $58 \times 10^3$ and $136\times 10^3$ MCS, respectively.
Thus it seems 
to  increase in a non-linear way. Also from Figure \ref{round_trip_time_not_normalized}, we can see that with increasing number of replicas, the round-trip time
is also increased.   This is because a single replica has to take a longer  path as $M$ is  increased. For example, for $M = 10$ it is only $9$ steps in one direction, and another $9$ in the other  to complete  a single round-trip, whereas  for $M=50$ it is 98 steps in both directions. From this  plot it is not clear what  is the optimal $M$. 
Therefore we report  the round-trip time divided by $2(M-1)$ (Figure \ref{round_trip_time}) in order to get time needed to perform a single step in $T^*$ space. Figure \ref{round_trip_time} showes  a clear minimum of the MCS time at a given M, for each N. While for $N=25$ and $M=6$, it takes about $2100$ MCS to perform a single step in $T^*$ space, when we increase $M$ we reach a minimum for $M = 12$ with $1280$ MCS. For higher $M$ we observe an increase of time needed to perform a single step.
For $N=50$ the minimum number of replicas is $M=8$, and  in this case it takes $4150$ MCS per single step. Again, with increasing $M$, the
number of MCS decreases, and for $M=20$ it reaches a minimum with $2370$ MCS. For $N=75$, we observe a similar behavior with minimum for $M=30$ at $3930$ MCS.
By dividing the minimum MCS by chain length, $N$,   we obtain  about $50$ (Figure \ref{no_of_steps_per_segment}). 


It is known that number of replicas grows like $\sqrt N$ where $N$ is a size of a system \cite{kofke, earl}. In principle, if one can find the optimal number of replicas for specified $N$, and then one can estimate the  optimal $M$ for any $N$. However,  in this  simulations it is not confirmed. If we  take $N=25$ with optimal $M=12$,  then optimal $M$ for $N=50$ and $75$ should be $17$ and $21$, respectively. However, it is known that with increasing polymer chain length, coil-to-globule transition moves to higher temperatures, so it has different specific heat which affects  the optimal $M$. Another explanation could be that our choice for definition of optimal $M$ is not the same as in the other studies \cite{rathore_earl}. On the other hand, we have only $3$ chain lengths so it is quite difficult to verify this scaling with only $3$ points. In our simulations we get an approximate ``scaling'':
\begin{equation}
M \sim N
\end{equation}
Average  exchange  probability between adjacent replicas for chain lengths $N= 25$, $ 50$ and $75$ is  $64\%$, $70\%$ and $74\%$, respectively. This is quite large compared to results obtained by Kone and Kofke \cite{kone_and_kofke} and by Rathore \textit{et al.} \cite{rathore_earl} for systems with constant specific heat. They found that optimal exchange probability is about $23\%$. 

\section{Conclusions}
We perform Monte Carlo simulations of homopolymer chains of lengths $N=25$, $50$ and $75$ in temperatures range $T_1^* = 1$ to $T_M^*=15$ and find the the  optimal number of replicas, $M$, based on the shortest time needed to perform a single step in the temperature space. We also find that for each chain length, $N=25,50$ and 75, the optimal $M$ is  $12, 20$ and $30$, respectively. The optimal $M$  is found to be approximately proportional to $M$, $M \sim N$ relation. 

However,  if one changes temperature range, then a set of the FOPT simulations should be run again to find optimal $M$ for this system. 
For  example, in this work we obtain that the  optimal $M=12$ for $N=25$, but if we  set $T_1^*=0.5$ and $T_M^*=10$, we would get different optimal $M$. Finding the optimal $M$ for different $T^*$ ranges is not a subject of this work, but it may be crucial. Another significant parameter which is not changed during the simulation is the replica exchange interval, $S$.

The FOPT is a promissing method for  simulating  polymer systems. This algorithm provides an iterative way to find optimal number of replicas, but it is quite CPU-time consuming (one has to perform numerous simulations to prepare an optimized $T^*$ set in which the targeted system is sampled). It can also be unstable, and sometimes the generated temperatures can be so inappropriate that the free energy bottlenecks are formed, which are difficult to overcome by the replicas' walk in the $T^*$ space. 

\begin{acknowledgments}
We gratefully acknowledge the computational grant from the
Poznan Supercomputing and Networking Center (PCSS) and grant N202 287338 from Polish Ministry of Science and Higher Education.
\end{acknowledgments}

\newpage
\bibliography{parallel_tempering}

\newpage
{\bf FIGURE CAPTIONS}
\linebreak[1]

\begin{figure}
\includegraphics*[width=13cm]{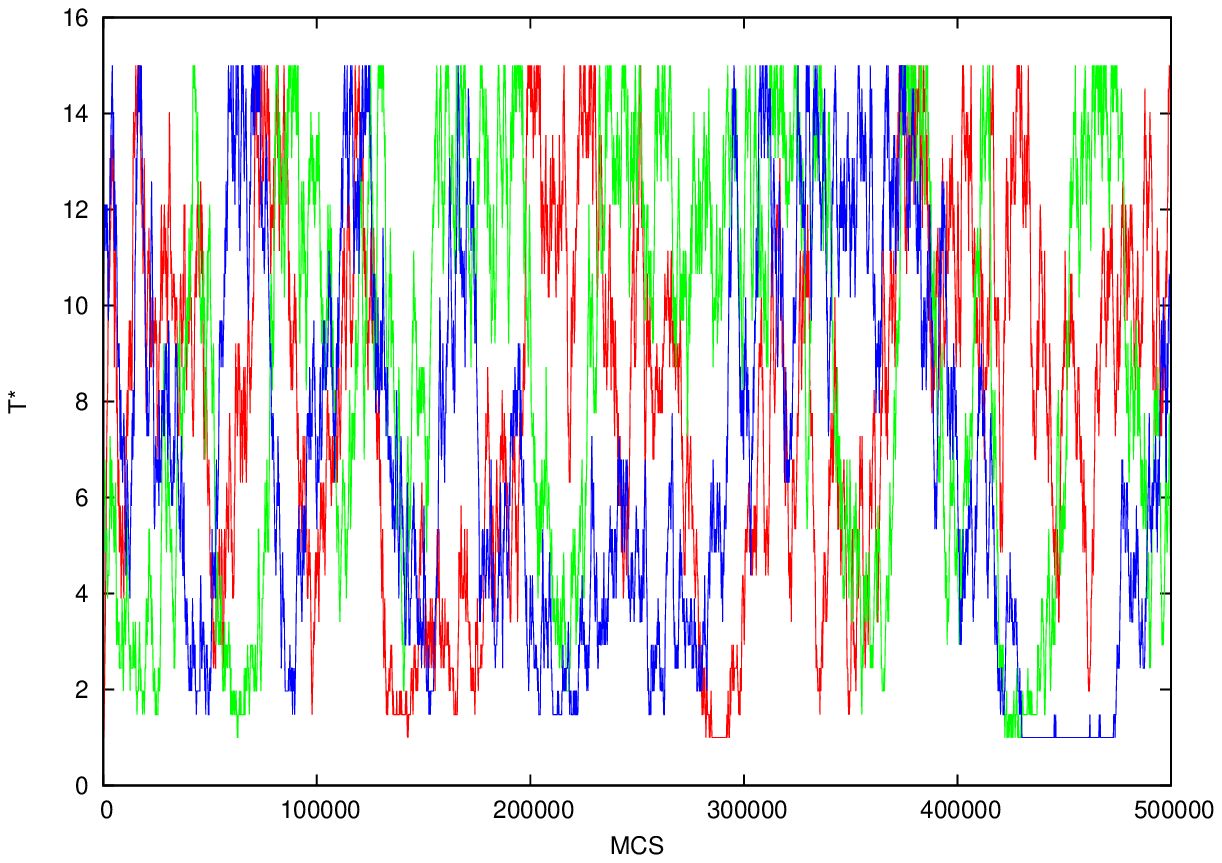}
\caption{\label{trace} Lewandowski et al.}
\end{figure}

FIG. {\ref{trace}:
Traces of three replicas in the temperature space for chain length, $N=75$, and the number of replicas, $M=30$, in the 1st round.
}

\begin{figure}
\includegraphics*[width=13cm]{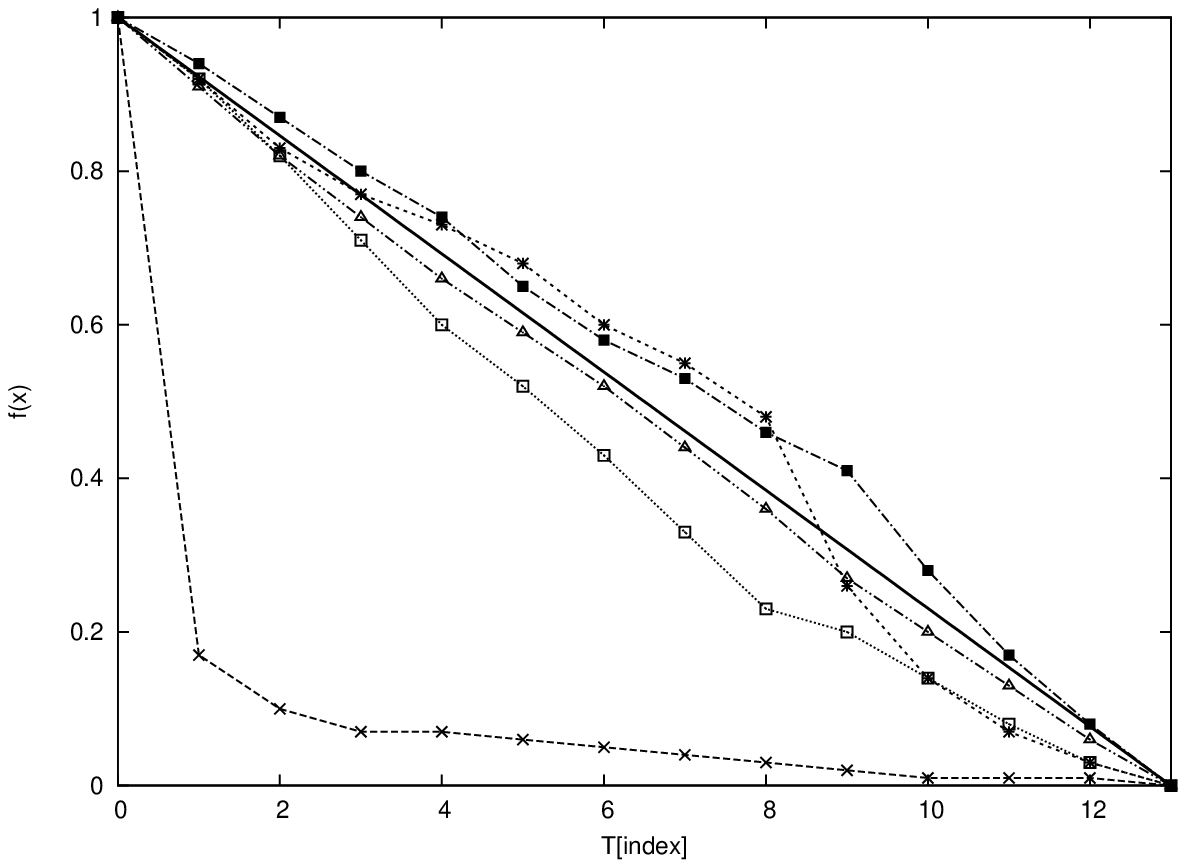}
\caption{\label{updown_histogram} Lewandowski et al.}
\end{figure}

FIG. {\ref{updown_histogram}:
The $f(T_i^*)$ histogram for each round of the FOPT simulation for $N=75$ and $M=14$: the 1st round (dashed line with crosses), the 2nd round (short dashed line with stars), the 3rd round (dotted line with squares), the 4th round (dash-dotted line with filled squares), the 5th round (dash-double-dotted line with triangles) and a reference (solid line).
}

\begin{figure}
\includegraphics*[width=13cm]{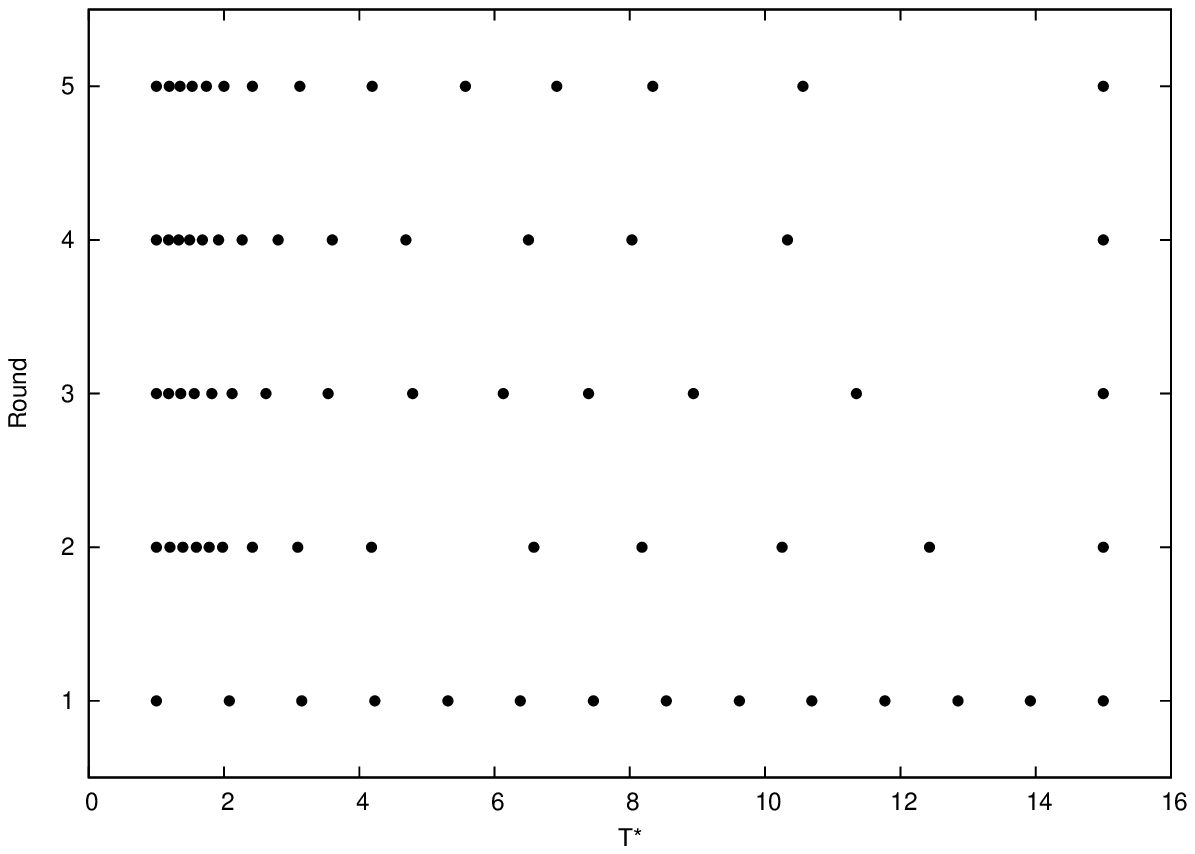}
\caption{\label{temperatures_set} Lewandowski et al.}
\end{figure}

FIG. {\ref{temperatures_set}:
The temperature set for each round of the  FOPT simulation for $N=75$ and $M=14$. In the first round we start with linear temperatures, and in next rounds temperatures are generated with the FOPT algorithm.
}

\begin{figure}
\includegraphics*[width=13cm]{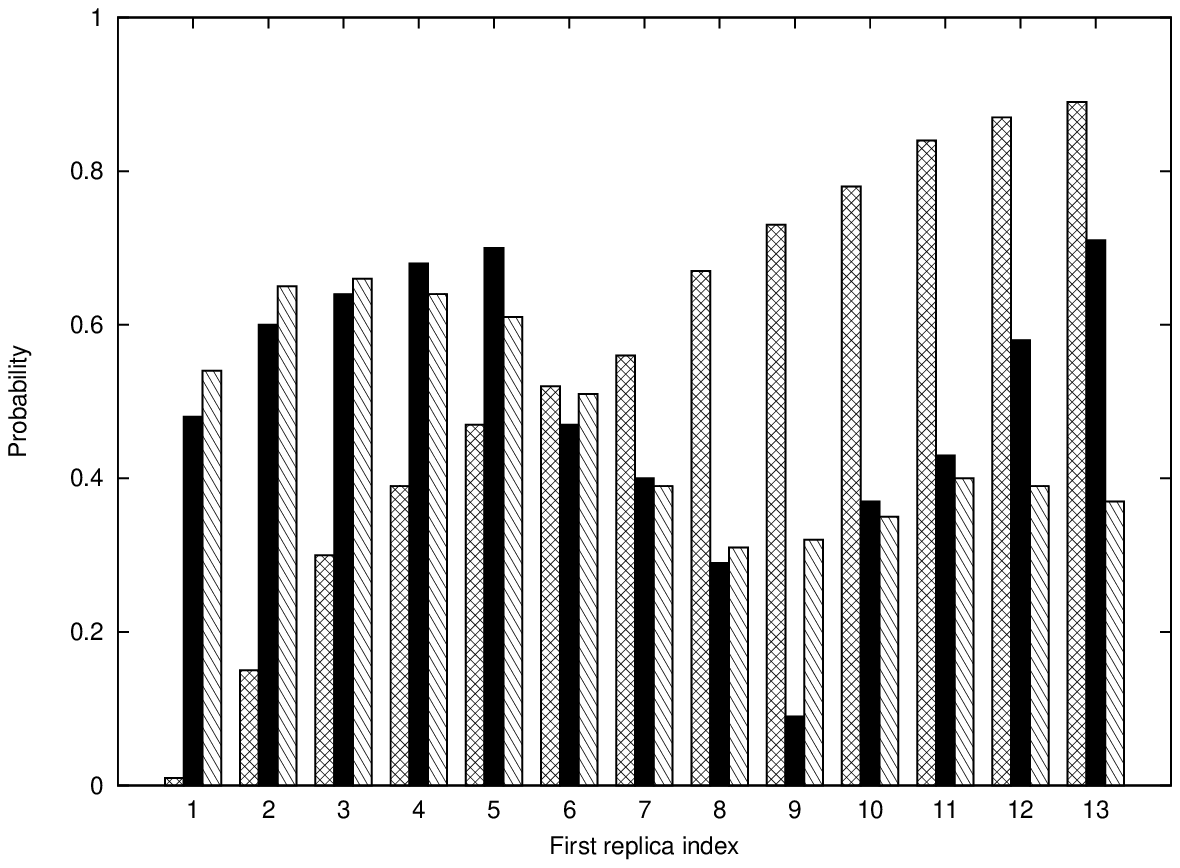}
\caption{\label{probability} Lewandowski et al.}
\end{figure}

FIG. {\ref{probability}:
Exchange probability for three FOPT rounds for $N=75$ and $M=14$. Boxes filled with crossed lines are for the 1st round, boxes filled with black are for the 2nd round and boxes filled with lines are for the 5th round.
}

\begin{figure}
\includegraphics*[width=13cm]{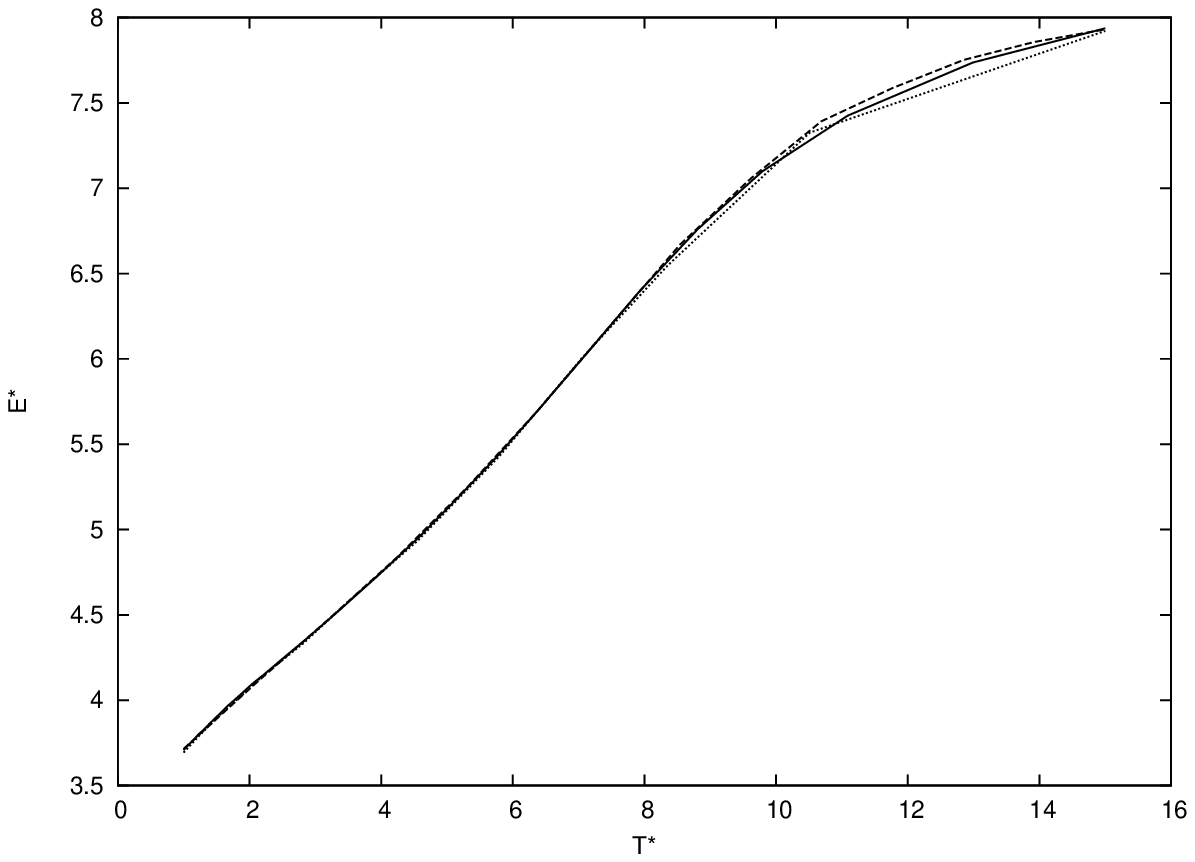}
\caption{\label{en} Lewandowski et al.}
\end{figure}

FIG. {\ref{en}:
Reduced energy per segment $E^*/N$ as a function of temperature $T^*$ for chain length $N=75$ with number of replicas $M=14$: the 1st FOPT round (dashed line), the 2nd round (dotted line) and a reference (solid line). For the reference we use $M=30$ and $2\times  10^7$ MCS.
}

\begin{figure}
\includegraphics*[width=13cm]{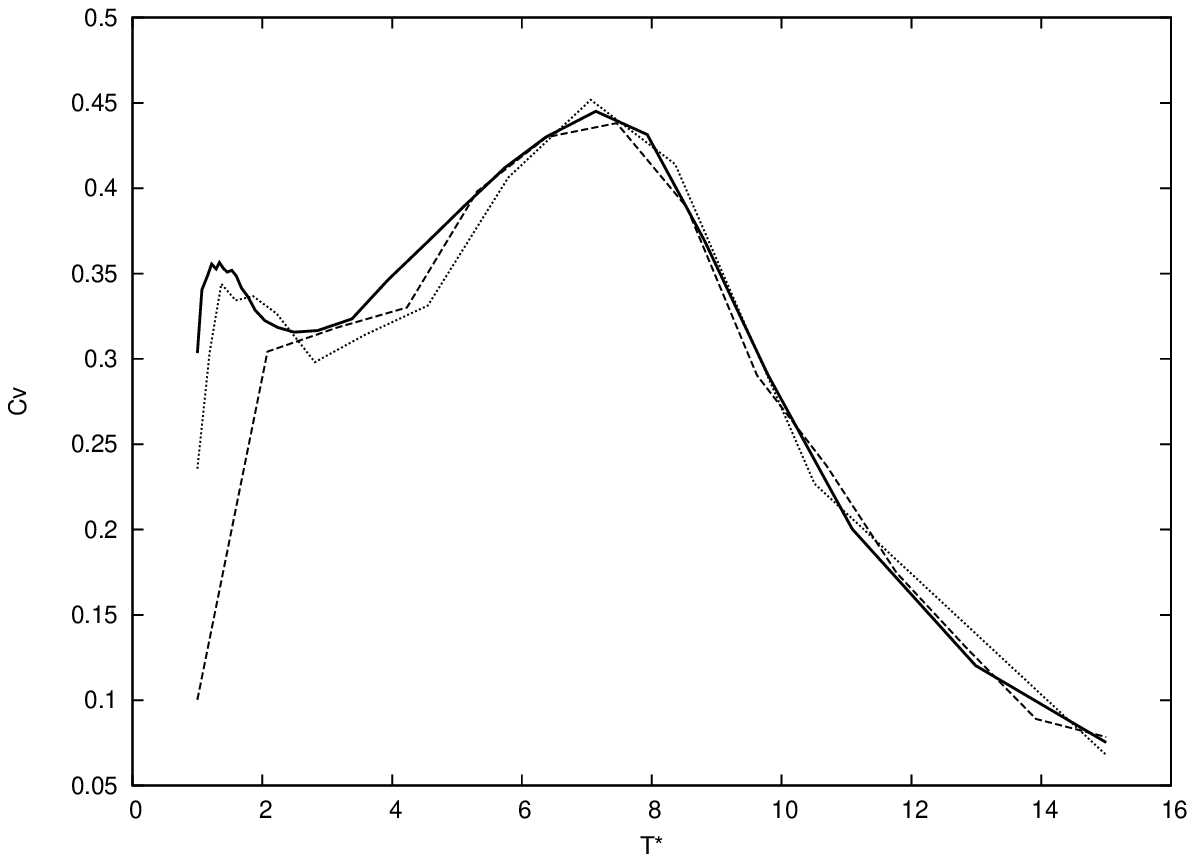}
\caption{\label{cv} Lewandowski et al.}
\end{figure}

FIG. {\ref{cv}:
Specific heat, $C_v$, as a function of temperature, $T^*$, for chain length, $N=75$, with the number of replicas, $M=14$: the 1st FOPT round (dashed line), the 2nd round (dotted line) and a reference (solid line). For the reference we use $M=30$ and $2\times 10^7$ MCS.
}

\begin{figure}
\includegraphics*[width=13cm]{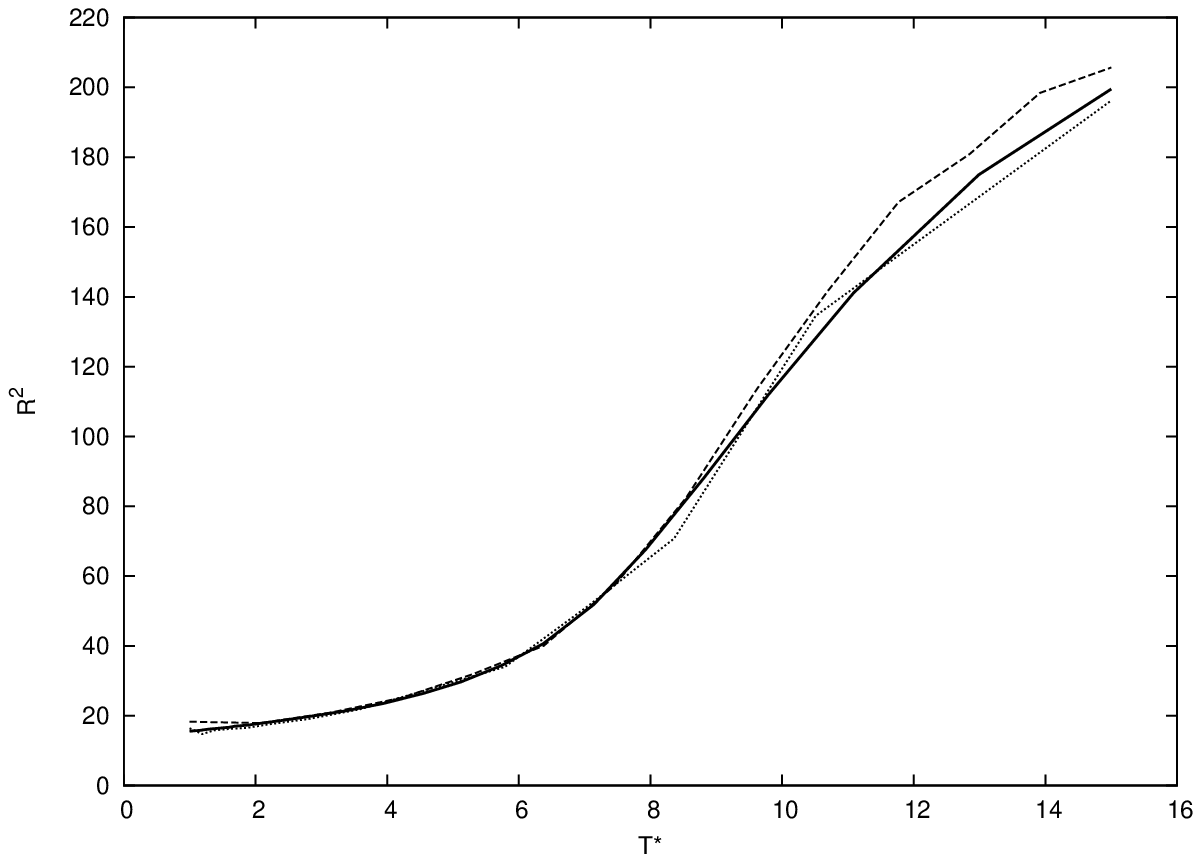}
\caption{\label{r2} Lewandowski et al.}
\end{figure}

FIG. {\ref{r2}:
Squared end-to-end distance as a function of temperature, $T^*$, for chain length $N=75$ with number of replicas $M=14$: the 1st FOPT round (dashed line), the 2nd round (dotted line) and a reference (solid line). For the reference we use $M=30$ and $2\times  10^7$ MCS.
}

\begin{figure}
\includegraphics*[width=9cm]{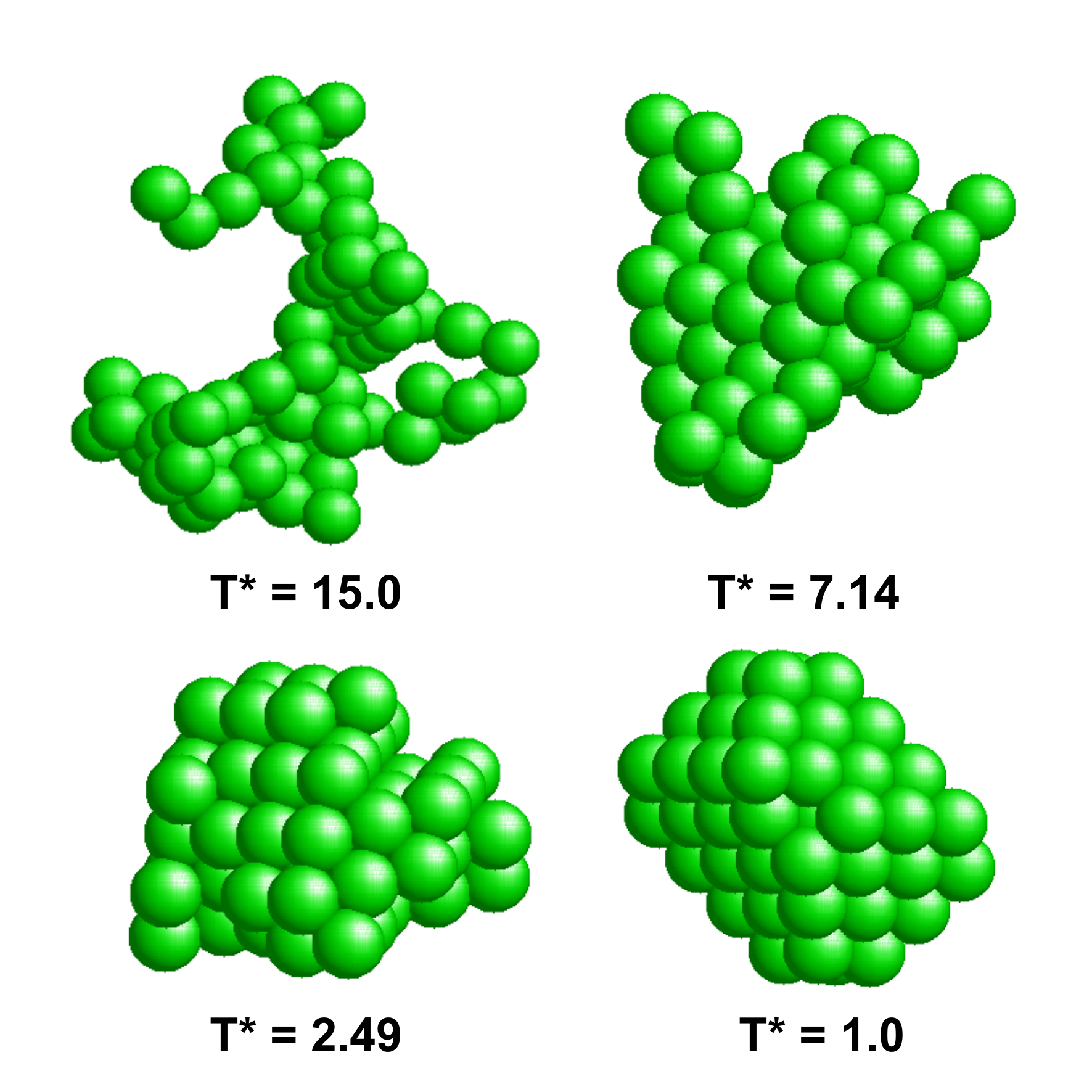}
\caption{\label{ordering} Lewandowski et al.}
\end{figure}

FIG. {\ref{ordering}:
Snapshots of structures of polymer chain of length $N=75$ for temperatures: $T^*=15$ (swollen state), $T^*=7.14$ (coil-to-globule transition), $T^*=2.49$ (globular state) and $T^*=1.0$ (globular state after another ordering transition in $T^* \approx 1.4$).
}

\begin{figure}
   \includegraphics*[width=13cm]{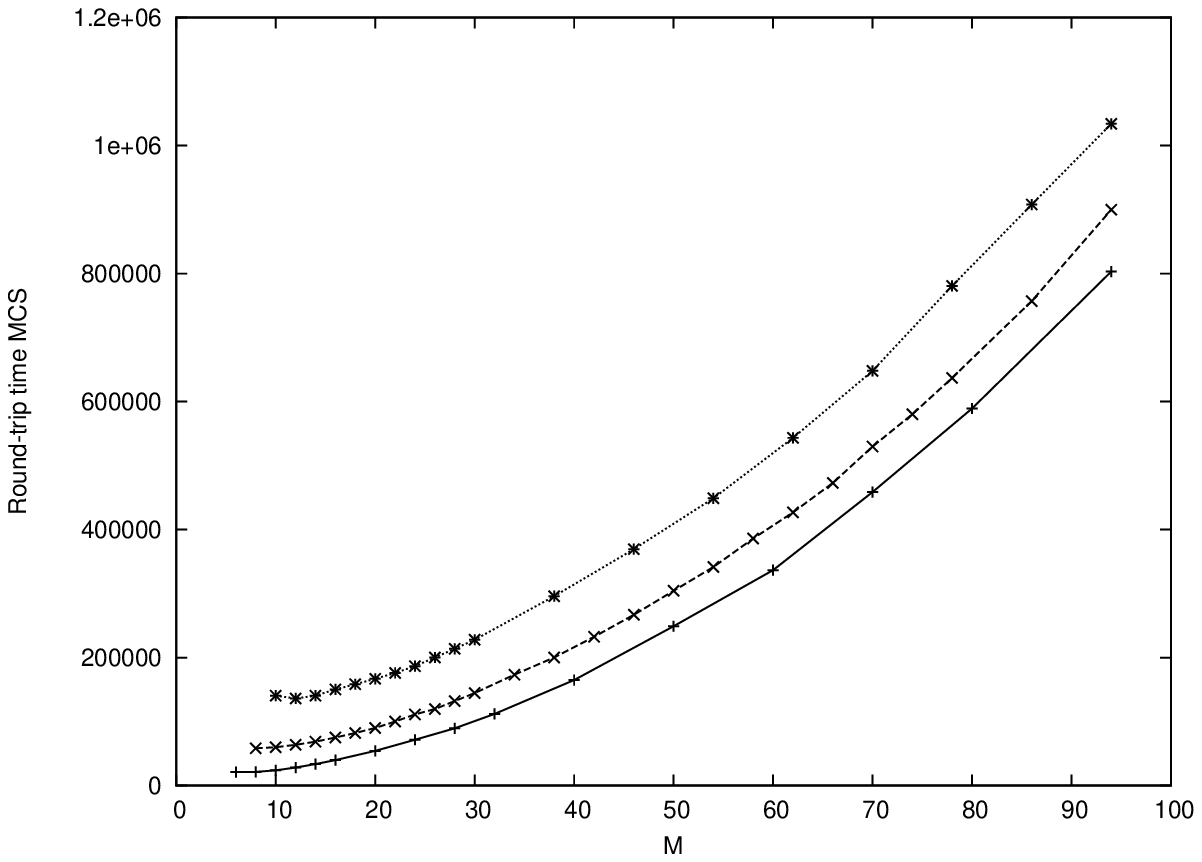} 	
 \caption{ \label{round_trip_time_not_normalized} Lewandowski et al. }
 \end{figure}

FIG. {\ref{round_trip_time_not_normalized}:
Round-trip time in MCS as a function of number of replicas $M$ for chain lengths: $N=25$ (solid line), $N=50$ (dashed line) and $N=75$ (dotted line).
}

\begin{figure}
   \includegraphics*[width=13cm]{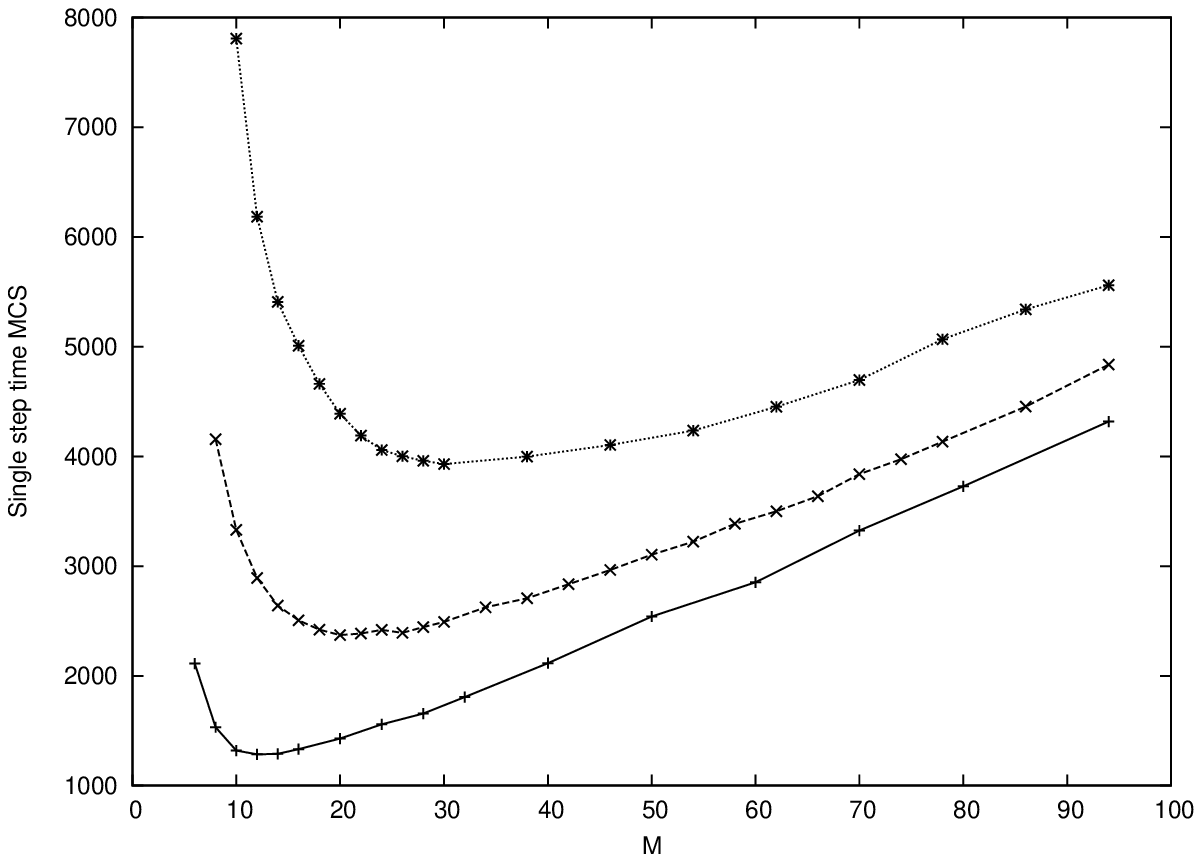}
 \caption{ \label{round_trip_time} Lewandowski et al. }
 \end{figure}

FIG. {\ref{round_trip_time}
Average time in MCS needed to perform a single step in temperature space as a function of number of replicas, $M$, for chain lengths: $N=25$ (solid line), $N=50$ (dashed line) and $N=75$ (dotted line).
}

\begin{figure}
   \includegraphics*[width=13cm]{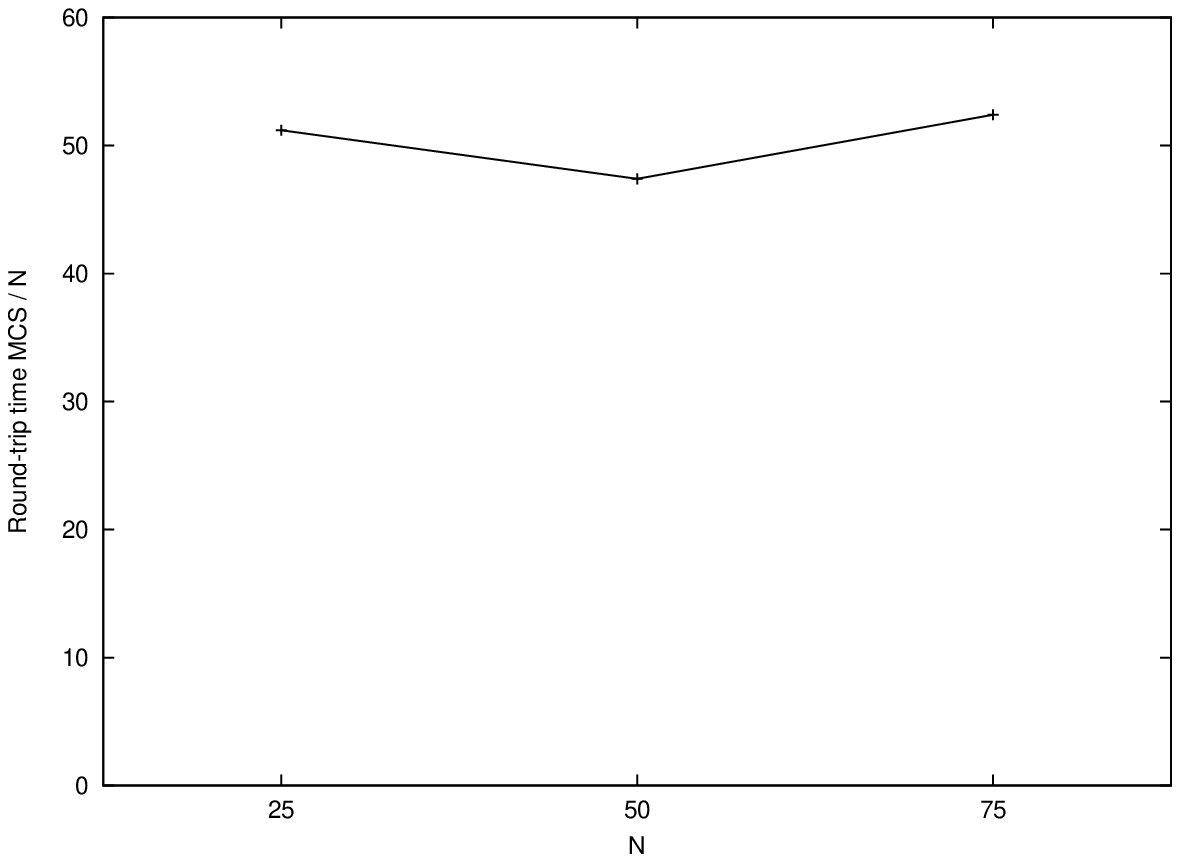}
 \caption{ \label{no_of_steps_per_segment} Lewandowski et al. }
 \end{figure}

FIG. {\ref{no_of_steps_per_segment}
Minimum average time needed to perform a single step in temperature space divided by chain length $N$ as a function of $N$ .
}

\newpage

\end{document}